# $Sb_2S_3$-Based Optical Switch Exploiting the Brewster Angle Phenomenon

**Diego Pérez-Francés,**[1,‡] **Gonzalo Santos,**[1,‡] **Josef Resl,**[2] **Maria Losurdo,**[3] **Yael Gutiérrez,**[3,4] **and Fernando Moreno**[1,*]

[1]*Department of Applied Physics Faculty of Sciences, University of Cantabria, 39005 Cantabria, Spain.*
[2]*Center for Surface and Nanoanalytics (ZONA), Johannes Kepler Universität, A-4040 Linz, Austria.*
[3]*CNR ICMATE, Corso Stati Uniti 4, I-35127 Padova, Italy.*
[4]*Physics Department, University of Oviedo, 33007 Oviedo, Spain.*
[‡]*Both authors contributed equally to this work.*
[*]*morenof@unican.es*

**Abstract:** Optical switches based on phase change materials (PCMs) hold great promise for various photonic applications such as telecommunications, data communication, optical interconnects, and signal processing. Their non-volatile nature as well as rapid switching speeds make them highly desirable for developing advanced and energy-efficient optical communication technologies. Ongoing research efforts in exploring new PCMs, optimizing device designs, and overcoming existing challenges are driving the development of innovative and high-performance optical switches for the next generation of photonics applications. In this study, we design and experimentally demonstrate a novel optical amplitude switch design incorporating PCM antimony trisulfide ($Sb_2S_3$) based on the Brewster angle phenomenon.

## 1. Introduction

Recent advancements in nanofabrication have led to significant technological breakthroughs in devices that rely on the interaction between light and nanostructured materials. [1, 2] As a result, nanophotonics has gained considerable attention across various fields, ranging from fundamental research to practical applications. [3–7] The increasing importance of high-speed data processing with minimal energy consumption has spurred the utilization of light for signal transmission and processing. [8] By employing nanostructures, the energy requirements and interaction timescales can be greatly reduced, enabling the integration of multiple devices that possess the ability to control, modulate, and process light signals with high speed and energy efficiency in the same chip. [9] The pursuit of more versatile nanophotonic devices has led to the exploration of phase-change materials (PCMs), offering the potential to create tunable and switchable nanophotonic systems through the modulation of their optical properties via external stimuli. [10–12] Reconfigurability offers significant advantages over conventional fixed-function devices by allowing dynamic adjustments in real-time, thus meeting the evolving demands of many photonic applications such as reconfigurable photodetectors, [13] reflective pixels, [14] integrated phase modulators, [15] tunable metasurfaces, [16–18] optical memories, [19] and in particular of optical switches. [20–22]

So far, reported optical switches are based on PCMs rely on $Ge_2Sb_2Te_5$ (GST). GST is a consolidated technology for rewritable optical disk storage and non-volatile electronic memories due to its high thermal stability, fast switching speed, and ability to undergo numerous rewriting cycles. [12,23,24] Most of GST based switches rely on absorption modulation because of the huge contrast in the extinction coefficient $k$ at 1550 nm between crystalline ($k$≈1.49) and amorphous ($k$≈0.12) phases. However, the high optical losses in GST make this material impractical for large-scale photonic integrated circuits (PICs) where light is guided through numerous phase change photonic routers. [25] Additionally, the optical losses of GST limit its exploitation for applications where phase modulation is required without affecting the amplitude of the signal. Moreover, because of the increasing extinction coefficient $k$ at shorter wavelengths for both phases, the integration of GST in photonic devices such as tunable metasurfaces operating in the visible or near infrared (near IR) becomes impractical. [16, 17]

In order to overcome this limitation, wide-bandgap PCMs are currently being investigated. These novel PCMs provides low-loss or even zero loss operation from the visible to mid IR range. [15, 25] $Sb_2S_3$ has been recently proposed as low-loss PCM. [26] In its crystalline form, $Sb_2S_3$ has a band gap of 1.6 eV, while the amorphous phase has the absorption edge at 2.2 eV, enabling a low-losses spectra region (extinction coefficient $k = 0$) extending from 1.6 eV to the IR with a modulation in the refractive index $\Delta n \approx 0.5$. [27] $Sb_2S_3$ has already been exploited in the design of reconfigurable structural color platforms, [28] as well as amplitude and phase light modulation in all-dielectric metasurfaces [16, 17] and Mach Zendher interferometers (MZIs) operating in the C- and O- communication bands. [29]

In this work, amorphous and crystalline samples of $Sb_2S_3$ are optically and structurally characterized to explore their potential in the design of an optical amplitude switch based on the Brewster angle phenomenon. [30] The Brewster angle corresponds to the angle where light with a specific polarization (polarization parallel to the plane of incidence, i.e., $p$-polarization) is completely transmitted through a transparent interface without any reflection. The value of the Brewster angle depends on the refractive index of both materials at the reflective interface. Thus, the Brewster angle is strongly affected by the phase of the PCM. When the non-reflection condition (i.e., Brewster angle) is met for one of the $Sb_2S_3$ phases, upon phase-change the Brewster condition is broken, leading to light reflection at the interface, and producing an optical contrast in reflection that can be harnessed for optical switching applications. $Sb_2S_3$ is the alternative low-loss PCM with larger transparency window. i.e, spectral range beyond which neither of the material's phases experience optical losses. This is really important for the optical switch based on Brewster angle, as this phenomenon only takes place if the material is transparent to light (neglectable extinction coefficient $k\approx0$). Under this optical switch configuration, we experimentally demonstrate an optical contrast of 22 dB. Noteworthy, we theoretically predict that this contrast could be further enhanced by tailoring the incident angle and the wavelength of the beam. These findings highlight the potential of using the Brewster angle effect in developing efficient optical switch devices based on low-loss wide band gap PCMs such as $Sb_2S_3$.

## 2. Methods

### 2.1. Sample fabrication

$Sb_2S_3$ films were deposited by chemical bath deposition (CBD) at 15°C using antimony chloride ($SbCl_3$) and sodium thiosulfate ($Na_2S_2O_3$) as precursors. At that temperature, $SbCl_3$ was dissolved in 2.5 g of acetone and 16.2 g of $Na_2S_2O_3$ was dissolved in 100 mL of deionized (DI) water. These two solutions were mixed, and then DI water was added to obtain a total volume of 178 ml of the liquid phase. The final mixture was vigorously stirred for 15 minutes. The glass substrates were cleaned in 5% NaOH at 90°C followed by 1N HCl, and absolute ethanol. After rinsing in DI water, the glass slides were dried at 80°C and then immersed in a vertical position in a beaker. The most important aspect for controlling the thickness in CBD is the deposition time. The duration the substrate is immersed in the deposition bath directly impacts the thickness. Longer deposition times generally result in thicker films, while shorter times produce thinner ones. Timing is essential for achieving the desired thickness. The $Sb_2S_3$ deposition time was between 2 - 4 h to have a film thickness in the range 100 nm to 1 $\mu$m.

## 2.2. Optical characterization

The optical properties of $Sb_2S_3$ thin films, namely the pseudodielectric function $\epsilon = \langle\epsilon_1\rangle + i\langle\epsilon_2\rangle$ = $(\langle n\rangle + i\langle k\rangle)^2$, being $n$ and $k$ the refractive index and extinction coefficient respectively, were measured from 0.75 (1650 nm) to 6.00 eV (205 nm) at room temperature using spectroscopic ellipsometry (Woollam 2000) operating at an angle of incidence varying from 55 to 75°.

Annealing of the as-deposited $Sb_2S_3$ was performed inside an enclosed heating cell for ellipsometry (Linkam THMSEL600). [31] The sample was initially placed on the heating block and ellipsometric measurements were taken at room temperature at 70°. The cell was then sealed, and a purging process with Ar was performed to remove oxygen and prevent oxidation. The sample was subsequently heated at a rate of 25 °C/min until it reached 265°C, inducing crystallization. The temperature of 265°C was maintained for at least 35 minutes to ensure stability in the ellipsometric readings, indicating the completion of the crystallization process. Finally, the temperature was gradually decreased at the same rate using $N_2$ cooling.

## 2.3. Strcutural characterization

Films, both as deposited and after an annealing cycle, were structurally characterized by Raman spectroscopy. The characterization was performed with a Horiba LabRam Aramis VIS×100 microscope objective (NA = 0.90) and excitation wavelength of $\lambda$ = 532 nm. Raman spectroscopy measurements were performed to address the crystallinity of the CBD $Sb_2S_3$ before and after annealing.

## 2.4. Reflectance measurements

### 2.4.1. Spectral

The optical system used for the spectral reflectance measurements is composed of a motorized Nikon Eclipse (ME600D) microscope with a ×10 objective (NA = 0.30), an Ocrean Optics USB200 spectrograph, and a spectral source OSL2 Fiber Illuminator. The reflectance spectra were processed and analyzed using in-house software fixing an integration time of 300 ms and an average scan of 30 cycles.

### 2.4.2. Angular

For the angular reflectance measurements, the sample was placed on a rotating platform. The incident beam, coming from a polarized Helium-neon laser ($\lambda$ = 633 nm), was reflected on the sample and then collected in a photodetector (NewPort-MODEL 818) connected to an optical power meter (NewPort-MODEL 1830-C).

## 2.5. Reflectance measurements

The reflectance of multilayer structures was calculated using the Transfer Matrix Method (TMM). [32] TMM allows to calculate the reflectance spectrum of an arbitrary system of homogeneous and non-magnetic multilayers by establishing conditions for the electric field along the boundary of two consecutive media. In the present case, the reflectance spectra of the fabricated system were calculated using the TMM in a multilayer configuration that mimics the experimental configuration (air/ $Sb_2S_3$/$SiO_2$). To take into account the effect of the microscope objective in the spectral reflectance measurement set-up, it has been considered the sum of all the reflected light that arrives to the objective lens. It has been angularly modelled with a normal distribution of mean 0 and standard deviation $\theta = \sin^{-1}(NA)$ where NA is the numerical aperture microscope objective (NA = 0.30).

## 3. Results

Figure 1(a) shows the Raman spectrum of a typical as-grown $Sb_2S_3$ film. The Raman spectrum is dominated by two broad features at≈140 and 300 $cm^{-1}$ assigned to Sb-S and S = S vibrational modes. The Raman spectrum after the annealing process plotted in Fig. 1(b) shows a series of sharp peaks assigned to vibrational modes $A_g$ at 157, 191, 284 and 303 $cm^{-1}$, $B_{1g}$ at 238 $cm^{-1}$ and $B_{2g}$ at 125, 313 and 303 $cm^{-1}$.

Figure 1(c) shows the complex refractive index ($n + ik$) of both $a$-$Sb_2S_3$ and $c$-$Sb_2S_3$. The complex refractive indices of $a$-$Sb_2S_3$ and $c$-$Sb_2S_3$ were derived from ellipsometric analysis [33] using a three-media air/film/substrate model, where the $SiO_2$ (glass) substrate was experimentally measured prior deposition and, the $a$-$Sb_2S_3$ and $c$-$Sb_2S_3$ layers were parameterized using the Cody–Lorentz oscillators model. [34] The optical band gap for each phase are at 2.1 eV and 1.8 eV respectively. Both values are in good agreement with other values reported in literature. [27] The thickness determined from the ellipsometric fitting were 290± 10 for $a$-$Sb_2S_3$ nm and 255 ± 10 nm for $c$-$Sb_2S_3$ . The discrepancy in thickness between both phases can be attributed to surface roughness and/or oxidation of the film surface. Figure 1(d) shows the refractive index contrast between $a$-$Sb_2S_3$ and $c$-$Sb_2S_3$. In general, a higher refractive index contrast results in the possibility of wider light amplitude/phase modulation. In the transparency region where $k_{am} = k_{cr} = 0$, the refractive index contrast is $\Delta n \approx 0.5$. This value is comparable to that measured by Delaney et al. in RF sputtered $Sb_2S_3$ films, i.e., $\Delta n \approx 0.6$. [15]

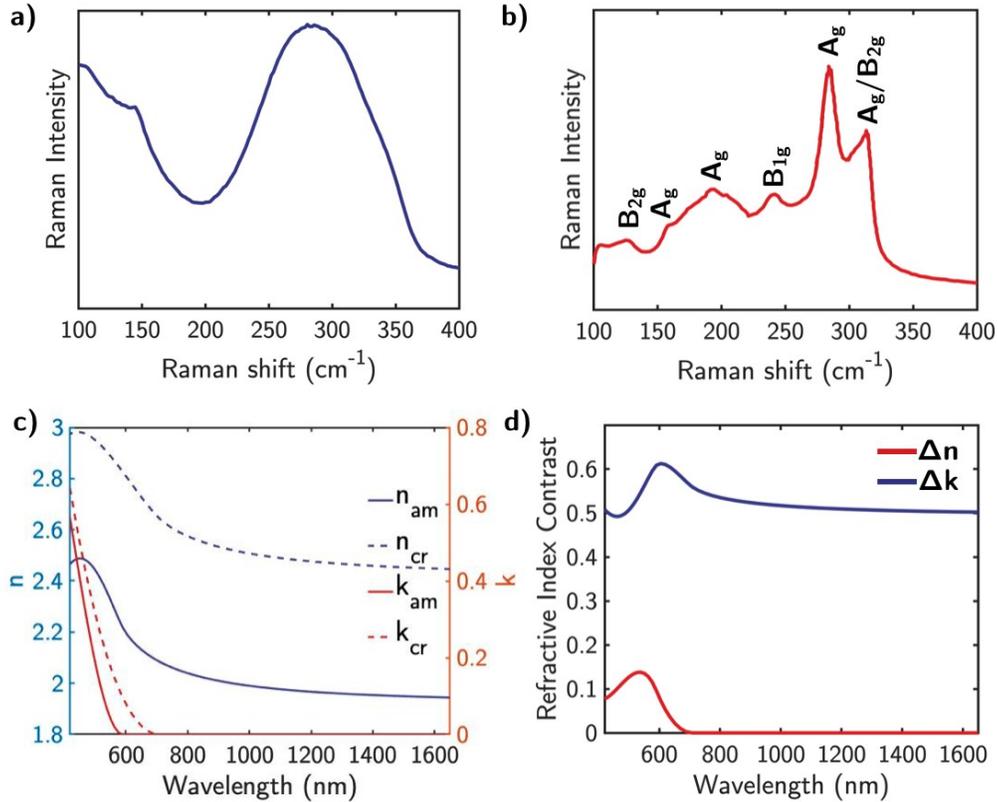

Fig. 1. Raman spectra of (a) $a$-$Sb_2S_3$ and (b) $c$-$Sb_2S_3$. (c) Complex refractive index, $n + ik$, of $a$-$Sb_2S_3$ and $c$-$Sb_2S_3$. (d) Refractive index contrast $\Delta n$ and $\Delta k$ between $a$-$Sb_2S_3$ and $c$-$Sb_2S_3$.

To create amplitude and phase modulators, the simplest configurations involves using a thin-film material with tunable complex permittivity. For this purpose, we examined a multilayer system designed to operate as an amplitude modulator in reflectance configuration consisting of a $Sb_2S_3$ film deposited on $SiO_2$ substrates.

In order to test the accuracy of the proposed theoretical model based in the TMM, reflectance measurements on the fabricated $a$-$Sb_2S_3$ and $c$-$Sb_2S_3$ films on $SiO_2$ were performed at normal incidence with unpolarized light and compared with equivalent simulated values. Figures 2(a,b) show the reflectance of the $a$-$Sb_2S_3$ and $c$-$Sb_2S_3$ films respectively. The black line represents the mean reflectivity, while the shadowed indicates the standard deviation. Individual reflectance measurements are displayed in Figure S1. The reflectance spectrum for $a$-$Sb_2S_3$ in Fig. 2(a) shows a main dip at $\approx$625 nm that, upon amorphous to crystalline phase-change, shifts to higher wavelengths ($\approx$675 nm, see Fig. 2(b)) due to the higher refractive index of the crystalline phase. The calculated reflectance for the $a$-$Sb_2S_3$ and $c$-$Sb_2S_3$ films on $SiO_2$ using the model described above are shown in Figs. 2(c,d). The black line represents the mean reflectance value and the shadowed region indicate the reflectance values covered by including the $\pm$10 nm error in the thickness determination of the films. The calculated reflectance spectra reproduce well the main features and their spectral shift upon amorphous-to-crystalline phase-change. Differences in the amplitude between the simulated and experimental reflectance spectra might be attributed to surface roughness not considered in the model.

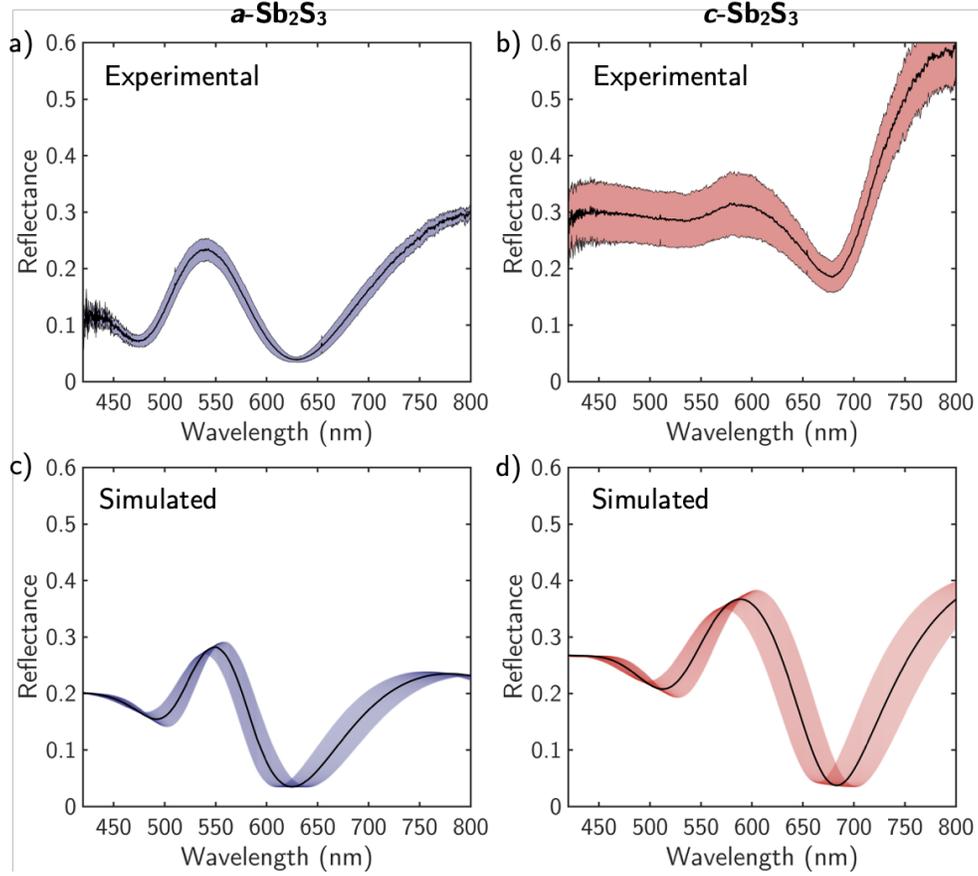

Fig. 2. Reflectance measurements at normal incidence for (a) $a$-Sb$_2$S$_3$ and (b) $c$-Sb$_2$S$_3$. The black line represents the mean reflectance, while the shadowed region indicates the standard deviation. Simulated reflectance for (c) $a$-Sb$_2$S$_3$ and (d) $c$-Sb$_2$S$_3$. The black line represents the mean reflectance value and the shadowed region indicate the reflectance values covered by including the ±10 nm error in the thickness determination of the films.

In this study, we propose a novel optical amplitude switch design to work in reflectance configuration incorporating PCM Sb$_2$S$_3$ and based on the Brewster angle phenomenon. Under the simplest approximation, i.e., considering air and the PCM Sb$_2$S$_3$ semi-infinite media, the Brewster angle $\theta_B$ is given by

$$\theta_B = \arctan(n_{PCM}/n_{air}) \quad (1)$$

where $n_{PCM}$ and $n_{air}$ are the refractive index of the phase change material and air, respectively.

The Brewster phenomenon on Sb$_2$S$_3$ films was investigated through reflectance measurements for fixed wavelength $\lambda$ = 633 nm using $p$-polarized light and modifying the angle of incidence. Figures 3(a,b) shows the angular dependent reflectance measurements taken on the $a$-Sb$_2$S$_3$ and $c$-Sb$_2$S$_3$ films respectively. The black line indicates the mean reflectivity values, while the shadowed region represents the standard deviation. Individual reflectance measurements are shown in Fig. S2. These measurements are supported by simulated angular dependent reflectance calculations under the same conditions (same angles of incidence and excitation wavelength) using the developed theoretical model based on TMM. For these simulations the dielectric functions and thicknesses have been extracted from experimental ellipsometric measurements. The calculated reflectance for $a$-Sb$_2$S$_3$ and $c$-Sb$_2$S$_3$ is shown in Figs. 3(c,d) respectively. The black line represents the mean reflectance value and the shadowed region indicate the reflectance values covered by including the ±10 nm error in the thickness determination of the films. Experimental and simulated values are in good agreement. The angle at which the reflectance reaches its minimum corresponds to the Brewster angle.

Figure 3e summarizes the value of the the Brewster angle for both $a$-Sb$_2$S$_3$ and $c$-Sb$_2$S$_3$ as determined in three different ways: (*i*) the Snell's law (equation (1)), (*ii*) from the minima in the experimental reflectance spectra, and (*iii*) from the minima in the simulated reflectance spectra. The Brewster angle for $a$-Sb$_2$S$_3$ cannot be directly obtained from the experimental data because of the absence of a well defined minimum in the reflectance curve (see Fig. 3(a)). This effect can be attributed to the presence of some residual $s$-polarized (i.e., polarization perpendicular to the plane of incidence) light in the beam during the experimental measurements (see Fig. S3).

The primary reason for the disparities in the Brewster angle values obtained through the TMM calculations and the Snell's law lies in the assumptions made during the derivation of equation (1). The latter relies solely on the refractive index of the materials at the interface, namely Sb$_2$S$_3$ and air. However, in the case of a stack of nanostructured materials as the one studied here (i.e., air/PCM/SiO$_2$), the Brewster angle is influenced by both the thicknesses and refractive index of the materials forming all the stack. These parameters govern the multiple reflections occurring at the various interfaces within the stack, leading to the observed variations. Therefore, as demonstrated in 3(e), the value of the Brewster angle obtained using the TMM, which consider the reflection at all the interfaces in the system, is in better agreement with the value from the experimental measurements.

Another consideration relies in the fact that the Brewster phenomenon is observed at interfaces between non-absorbing materials (i.e., extinction coefficient $k$ = 0). This condition is met when the photon energy of the incident beam lies below the absorption onset of both materials forming the interface. In contrast, if the one or both materials exhibit absorption (i.e., extinction coefficient $k \neq 0$), a similar effect can be observed know as pseudo-Brewster. [30] Under the pseudo-Brewster condition, the reflectance minimum increases deviating from absolute zero reflectance of the pure Brewster phenomenon. Understanding this distinction is crucial for optimizing optical devices based on this phenomenon and manipulating light properties under this condition.

In the reflectance values of Figs. 3(b,d), obtained using an incident wavelength of $\lambda$ = 633 nm, the refractive index of $c$-Sb$_2$S$_3$ shows non-zero extinction coefficient (i.e., complex refractive index equal to 2.750 +0.035$i$). Therefore, for $c$-Sb$_2$S$_3$, the pseudo-Brewster condition is met and a non-zero value of the reflectance minimum is obtained. Therefore, considering the absorption onsets of both $a$-Sb$_2$S$_3$ and $c$-Sb$_2$S$_3$ (i.e., 1.6 and 2.2 eV respectively), to meet the pure Brewster condition in both phases simultaneously, operating wavelengths higher than 688 nm are needed.

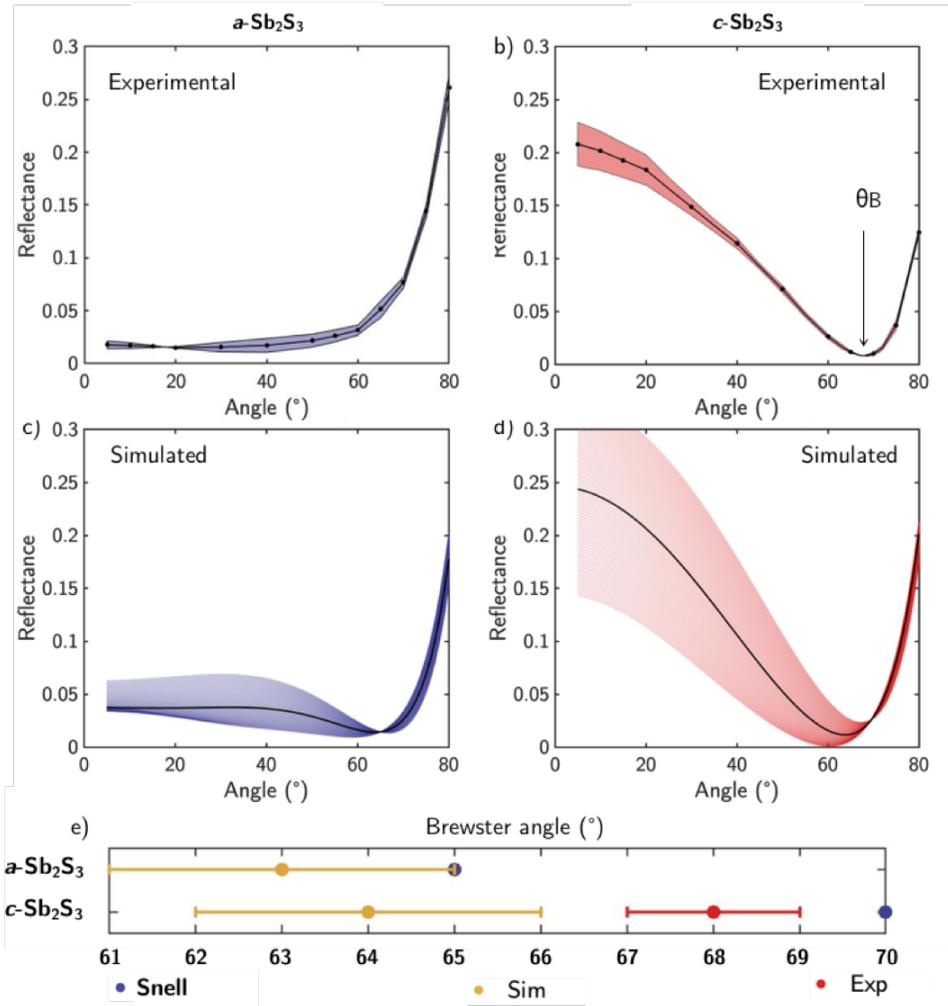

Fig. 3. Reflectance measurements at a wavelength of $\lambda$=633 nm modifying the incident angle for (a) $a$-Sb$_2$S$_3$ and (b) $c$-Sb$_2$S$_3$. The black line represents the mean reflectivity, and the shadowed region indicates the standard deviation. The black arrow pointing the minimum of the curve indicates the Brewster angle ($\theta_B$). Simulated reflectance at a wavelength of $\lambda$=633 nm modifying the incident angle for (c) $a$-Sb$_2$S$_3$ and $c$-Sb$_2$S$_3$ (d). The black line represents the mean reflectance value and the shadowed region indicate the reflectance values covered by including the $\pm$ 10 nm error in the thickness determination of the films.(e) Brewster angle for $a$-Sb$_2$S$_3$ and $c$-Sb$_2$S$_3$ considering Snell equation (blue), simulations based on the TMM (yellow) and experiment (red). The error estimated from the Brewster angle from the simulations and experimental data is 2° and 1° respectively. The Brewster angle obtained from equation (1) is considered without error.

Figures 4(a,b) show the simulated reflectance for $p$-polarized light as a function of the angle of incidence and wavelength for both air/$a$-Sb$_2$S$_3$/SiO$_2$ ($R_{am}$) and air/$c$-Sb$_2$S$_3$/SiO$_2$ ($R_{cr}$) systems studied in this work. The continuous and dashed black lines represent the minima in the reflectance spectra, i.e., the Brewster angles for $a$-Sb$_2$S$_3$ and $c$-Sb$_2$S$_3$ respectively. To evaluate the amplitude switching of the system upon amorphous-to-crystalline Sb$_2$S$_3$ phase-change, Fig.

4c shows the calculated optical contrast, $C$, in decibels (dBs) defined as

$$C = 10\log(R_{cr}/R_{am}) \qquad (2)$$

The highest optical contrasts (blue and red spots) match the maximums and minimums of the Brewster angle as a function of the wavelength. The highest absolute value of the calculated $C$ is 33 dBs. To demonstrate it, reflectance as a function of the wavelength has been measured for an angle of incidence of 55° (see white vertical line Fig. 4c). The experimental and simulated spectral reflectance values are shown in Figure S4. Figure 4d compares the simulated and experimental $C$ spectra for an angle of incidence 55°. Interestingly, this angle matches the Brewster angle of air/$a$-Sb$_2$S$_3$/SiO$_2$ at 1045 nm, and of air/$c$-Sb$_2$S$_3$/SiO$_2$ at 660 and 1190 nm. Therefore, for this specific angle incidence, $C$ present three maximum values as shown Fig. 4d. Both experimental and simulated line shapes of the $C$ spectra are in good agreement with small shifts of the minima/maxima. Nevertheless, experimental values of the maxima of $C$ are lower as compared with the calculated ones, probably due to undesired effects such as surface roughness or residual $s$-polarized light in the incident beam. The maximum value of $C$ experimentally measured is $C = 22$ dB at 1080 nm as opposed to the theoretically predicted value of $C = 33$ dBs at 1045 nm. The values of $C$ reported here under the Brewster condition underscore the capacity to harness the Brewster angle effect in the development of high $C$ optical switches.

For a practical implementation of the present proposed optical switch, a light guide should be designed on top of the PCM in such a way that the light entering the switching system hits the surface separation guide-PCM, at the Brewster condition for one of the PCM phases (OFF state) and not for the other (ON state). The reflected energy is in turn symmetrically collected to continue its propagation along the guide. Under this configuration, the switching condition can be even improved by using a symmetrical arrangement in such a way that the guide and the substrate are manufactured with the same material. It is important to highlight that the reflected electromagnetic energy for the ON state is low, of the order of 1% of the incident one, but this is not a serious drawback for the switching purposes given the high values of $C$ obtained. Different PCMs are suitable for this application. The choice of PCM is directly linked to the specific region of the spectrum in which the optical switch operates. As previously mentioned, it's important to highlight that the Brewster phenomenon exclusively emerges when the material presents no losses, precisely when it falls within the optical transparency window of the PCM.

For completeness and as an example, Figs.4(e,f) shows the simulated optical contrast in dBs, considering $p$-polarization, as a function of angle of incidence and wavelength for amorphous and crystalline samples while maintaining an equal thickness for both phases (255 nm and 290 nm, respectively). Ensuring equal film thickness in both phases upon phase-change promotes $C$. For instance, at an angle of incidence of 55°, the maximum value of $C$ experimentally measured of 22 dB at 1080 nm and theoretically predicted to be $C = 33$ dBs at 1045 nm, is increased to $C = 38$ dBs at 930 and 1050 nm by keeping the thickness of the layer to either 255 nm or 290 nm respectively.

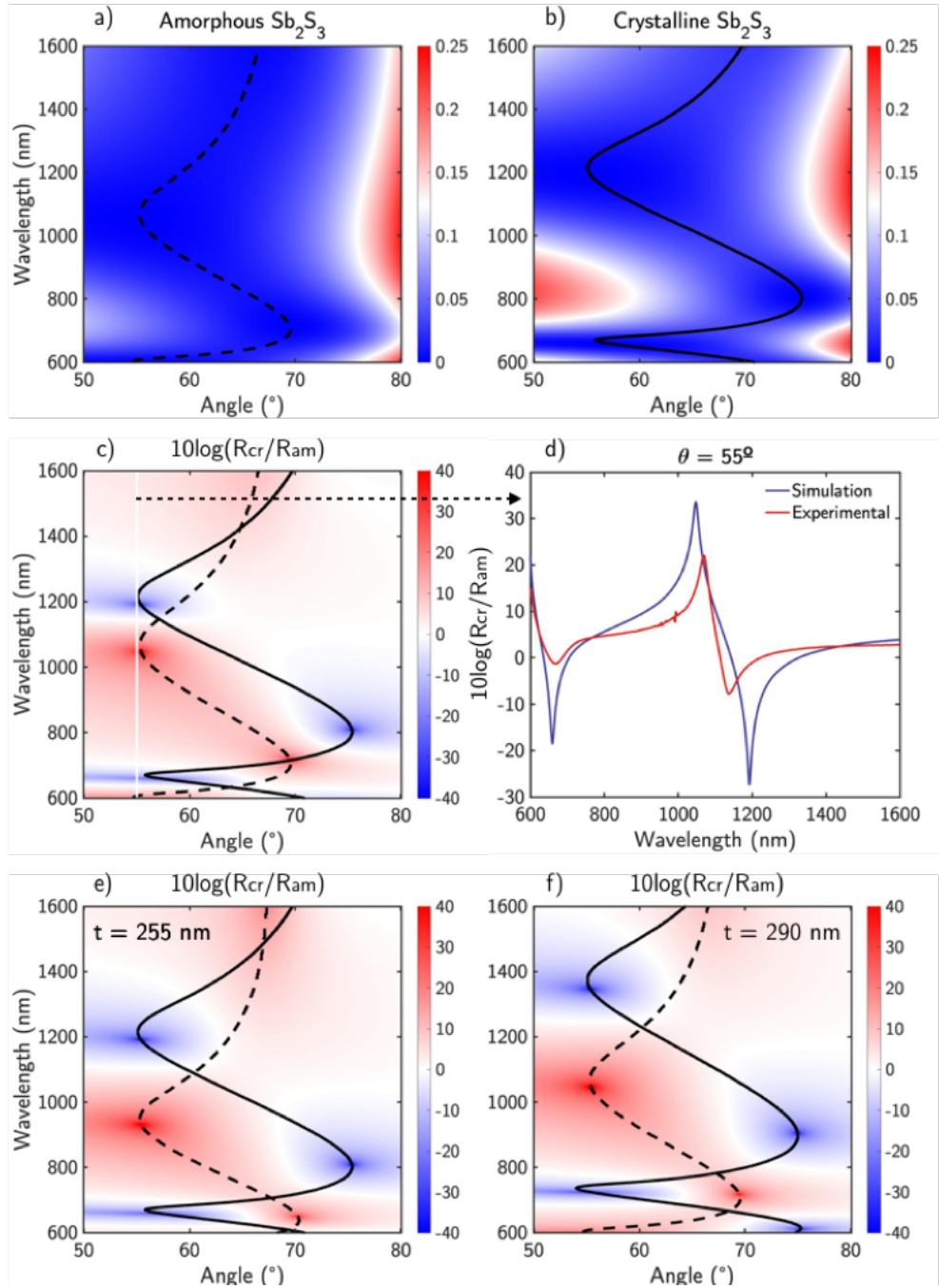

Fig. 4. Simulated reflectance for *p*-polarized light as a function of the angle of incidence and wavelength for (a) air/*a*-Sb$_2$S$_3$/SiO$_2$ ($R_{am}$) and (b) air/*c*-Sb$_2$S$_3$/SiO$_2$ ($R_{cr}$). The continuous and dashed black lines represent the points of minimum reflectance in the spectrum, corresponding to the respective Brewster angles. (c) Simulated optical contrast in decibels (dBs) ($C = 10\log_{10}(R_{cr}/R_{am})$) for *p*-polarization as a function of the angle of incidence and wavelength. (d) Calculated and experimentally measured $C$ spectra in dBs for an incident angle of 55°. Simulated optical contrast in decibels (dBs), for *p*-polarization, as a function as a function of the angle of incidence and wavelength amorphous and crystalline samples with equal thickness of (e) 255 nm and (f) 290 nm for both phases.

## 4. Conclusions

In conclusion, we conducted optical and structural characterizations on both amorphous and crystalline samples of $Sb_2S_3$ with the aim of investigating the feasibility of designing an amplitude optical switch based on the Brewster angle phenomenon working on the reflectance configuration. The designed optical switch consist of a simple stack of PCM $Sb_2S_3$ film on a $SiO_2$ substrate. Under this configuration, we successfully experimentally demonstrated an optical contrast of 22 dB upon amorphous-to-crystalline phase change of the $Sb_2S_3$ film. To implement this configuration in an optical circuit, a light guide should be designed on top of the PCM in such a way that the light entering the switching system hits the surface separation, guide-PCM, under the Brewster condition for one of the PCM phases (OFF state) and not for the other (ON state). The contrast between both states can be even improved by using a symmetrical arrangement: guide and substrate made of the same material. These findings demonstrate the potential of leveraging the Brewster angle effect in the development of efficient optical switch devices, paving the way for novel advancements in the field of optoelectronics and nanophotonics for communications and optical memories applications.

**Funding.** European Union's Horizon 2020 research and innovation program (No 899598 – PHEMTRON-ICS)

**Acknowledgments.** The authors acknowledge Pedro Valle for reflectance measurements and helpful discussion.

**Disclosures.** The authors declare no conflicts of interest.

**Data availability.** Data underlying the results presented in this paper are not publicly available at this time but may be obtained from the authors upon reasonable request.

**Supplemental document.** See Supplement 1 for supporting content.